\DocumentMetadata{%
	pdfstandard=A-3u,
	pdfversion=1.7,
	lang=en-US,
}

\documentclass[nofoot,balance,colorlinks,upint,subscriptcorrection,varvw,mathalfa=cal=boondoxo,spanish,german,vietnamese,russian,greek]{asmeconf}

\hypersetup{%
	pdfauthor={John H. Lienhard},									  
	pdftitle={ASME Conference Paper LaTeX Template},                  
	pdfkeywords={ASME conference paper, LaTeX template, BibTeX style},
	pdfsubject = {Describes the asmeconf LaTeX template},			  
	pdflicenseurl={https://ctan.org/pkg/asmeconf},
}

\usepackage{amsmath}
\usepackage{subcaption}
\usepackage{algorithm}
\usepackage{algorithmic}

\begin{document}




\title{Generative Multimodal Multiscale Data Fusion for Digital Twins in Aerosol Jet Electronics Printing
} 
 
%
%
%

\SetAuthors{%
    Fatemeh Elhambakhsh\affil{1}\JointFirstAuthor, 
	Suk Ki Lee\affil{2}\JointFirstAuthor,  
	Hyunwoong Ko\affil{2}\CorrespondingAuthor{Hyunwoong.Ko@asu.edu}
	}

\SetAffiliation{1}{School of Computing and Augmented Intelligence, Arizona State University, Tempe, AZ}
\SetAffiliation{2}{School of Manufacturing Systems and Networks, Arizona State University, Mesa, AZ}


\maketitle





\begin{abstract}
The rising demand for high-value electronics necessitates advanced manufacturing techniques capable of meeting stringent specifications for precise, complex, and compact devices, driving the shift toward innovative additive manufacturing (AM) solutions. Aerosol Jet Printing (AJP) is a versatile AM technique that utilizes aerosolized functional materials to accurately print intricate patterns onto diverse substrates. Machine learning (ML)-based Process-Structure-Property (PSP) modeling is essential for enhancing AJP manufacturing, as it quantitatively connects process parameters, structural features, and resulting material properties. However, current ML approaches for modeling PSP relationships in AJP face significant limitations in handling multimodal and multiscale data, underscoring a critical need for generative methods capable of comprehensive analysis through multimodal and multiscale fusion. To address this challenge, this study introduces a novel generative modeling methodology leveraging diffusion models for PSP data fusion in AJP. The proposed method integrates multimodal, multiscale PSP features in two phases: (1) registering the features, and (2) fusing them to generate causal relationships between PSP attributes. A case study demonstrates the registration and fusion of optical microscopy (OM) images and confocal profilometry (CP) data from AJP, along with the fine-tuning of the fusion step. The results effectively capture complex PSP relationships, offering deeper insights into digital twins of dynamic manufacturing systems.
\end{abstract}







\section{Introduction}
The demand for high-value electronics has surged across sectors such as consumer electronics, automotive, healthcare, and aerospace \cite{won2023transparent, marchio2014automotive}. 
These advanced devices require increasingly complex components that support precise material processing, energy efficiency, and sustainability, as well as sophisticated product design, process monitoring, control, and qualification \cite{espera20193d}. 
Manufacturing these components demands high-precision techniques to meet stringent specifications for geometry, material density, and electrical properties \cite{alhendi2022printed}. 
As devices become more compact, multifunctional, and even personalized, achieving meticulous control over these structural and material properties at the microscale is essential \cite{chen2024customized}. 
Traditional manufacturing methods often fall short of meeting these requirements, pushing the industry toward innovative additive manufacturing (AM) solutions \cite{lu2018additive}.

Aerosol Jet Printing (AJP) has emerged as a powerful AM method, particularly valued for its precision and versatility \cite{wilkinson2019review, guo2024applications}. 
As a type of direct writing technique, AJP enables the manufacturing of fine, complex electronic patterns using a variety of functional materials, including metals, semiconductors, polymers, and biological materials.
It can be applied on diverse substrates, whether planar or highly complex three-dimensional (3D) surfaces, without limitations imposed by surface morphology \cite{gibson2021additive, guo2024applications}. 
The process relies on either ultrasonic or pneumatic atomization to generate an aerosol, which is then focused into a stream of fine droplets that are precisely deposited onto the substrate. 
This flexibility allows AJP to be applied in various digital manufacturing applications, including the fabrication of electronic circuits, flexible displays, biosensors, and semiconductor packaging \cite{wilkinson2019review, guo2024applications}.

Achieving consistent quality and optimal performance in AJP manufacturing requires an in-depth understanding and sophisticated control of complex Process-Structure-Property (PSP) relationships \cite{ko2021machine, ko2023framework, wilkinson2019review, secor2018principles, lee2024amtransformer}. 
Machine learning (ML)-based PSP modeling plays a crucial role in improving AJP manufacturing by establishing quantitative relationships between process parameters, such as gas flow rates and printing speed, structural features, such as line morphology and thickness, and the properties of manufactured parts, such as electrical conductivity and resistance \cite{secor2018principles,lombardi2019image,meenakshisundaram2021insights}.
Leveraging ML’s abilities to learn complex, nonlinear, and hidden patterns, the ML-based modeling approaches establish quantitative relationships between process parameters, structural features, and properties of manufactured outcomes. 
ML-based PSP modeling aligns with broader trends in AM, where such approaches have demonstrated success in predicting and optimizing manufacturing outcomes \cite{yan2018integrated, yan2018modeling, ko2023framework, lee2024amtransformer}.

Despite advances in ML-based PSP modeling in AJP, existing approaches still face critical limitations in handling multimodal, multiscale manufacturing data fusion.
The spatiotemporal fusion of multimodal, multiscale manufacturing data is critical for developing Digital Twins (DTs) that offer a comprehensive, unified view of physical phenomena within complex manufacturing processes \cite{yang2024holistic, lu2024overarching}. 
By integrating PSP data across multiple modalities and scales, data fusion enables DTs to reveal hidden information about underlying process mechanisms and physical phenomena that might otherwise go undetected.
This holistic insight enhances the understanding of complex electronics AM systems, ultimately supporting the optimization of new product designs with a deeper knowledge of manufacturability. 

In electronics AM, PSP data fusion opens new opportunities for precision and control, unveiling novel PSP insights and advancing the creation of customized electronic components.
However, conventional ML models often focus on specific aspects of the PSP chain rather than providing an integrated spatiotemporal analysis of the PSP relationships. 
This fragmented approach could lead to an incomplete understanding of causal relationships and hinder accurate prediction. 
These limitations underscore the urgent need for a PSP fusion approach that can simultaneously analyze multimodal, multiscale data and generate their fusion while capturing complex dependencies, enabling the improvement of process control and quality assurance in AJP manufacturing.

To address this challenge, this study presents a generative modeling-based approach that utilizes diffusion modeling to fuse AJP PSP data. First, the methodology extracts features and spatially and temporally registers multimodal, multiscale PSP data to ensure that PSP features of interest are prepared for the fusion step. Second, using registered PSP data, the study fuses AJP PSP features with a fusion method based on a Denoising Diffusion Implicit Model (DDIM) \cite{song2020denoising, zhao2023ddfm}. 
The proposed data fusion methodology newly enables the
prediction of AJP PSP causality using multimodal, multiscale AJP
data and the generation of novel, synthesized PSP features
based on learned distributions. By leveraging the predictive
power of diffusion modeling, this research offers a more
nuanced and comprehensive understanding of complex AJP processes.
This, in turn, has the potential to significantly improve the way in which
to simulate, predict, and optimize AJP processes, ultimately
leading to advances in manufacturing quality, efficiency,
and innovation.

The remainder of this paper is organized as follows. Section 2 reviews the relevant literature on the AJP process, ML-based PSP modeling in AJP, and data fusion methods. Section 3 outlines the problem statement, and Section 4 introduces the proposed generative modeling-based PSP fusion framework. Section 5 presents experimental results demonstrating the effectiveness of our approach. Lastly, Section 6 offers concluding remarks and suggests future research directions.

\section{Literature Review}
\label{Lit}
An AJP process consists of five stages that uniquely contribute to the high-resolution and precise deposition of materials on diverse substrates \cite{secor2018principles}. 
The process begins with aerosol droplet generation (atomization), where either ultrasonic or pneumatic atomization creates fine droplets. Ultrasonic atomization uses high-frequency sound waves to break the ink into stable microscale droplets, while pneumatic atomization shears ink into polydisperse droplets through high-velocity gas flow, selectively directing droplets with higher inertia toward deposition. 
Following atomization, the aerosol transport stage moves the droplets through a carrier gas flow in the mist tube, where gravitational settling and wall adhesion pose potential transport losses. 
Next, in aerosol beam collimation, the aerosol stream is guided by a surrounding sheath gas, narrowing it into a concentrated beam and minimizing material accumulation on the walls. 
This collimation is followed by aerodynamic focusing, where the beam is further tightened by a converging nozzle that centers the droplets for enhanced resolution. 
Finally, in the impaction stage, the focused aerosol exits the nozzle and impacts the substrate, translating the accumulated control of prior stages into precise, high-quality printed patterns on a variety of substrates \cite{wilkinson2019review, secor2018principles, guo2024applications}.

Recent studies using ML-based modeling methods to reveal the complex PSP relationships in AJP has shown promising results. 
The study conducted by Tafoya and Secor investigated how printhead geometry affects the characteristics of printed lines \cite{tafoya2020understanding}. 
Their study enhanced the understanding of aerosol flow dynamics and drying effects, which are key factors affecting the resolution of printed lines across various nozzle diameters. 
The study utilizes a Support Vector Machine (SVM) framework. The framework analyzes the relationship between these factors and identifies the linear deposition rate as a critical variable impacting printing quality. 
ML for process-structure modeling approaches has also been extended to a knowledge transfer framework, as shown in a study by Zhang \textit{et al.} \cite{zhang2021knowledge}.
Their framework is intended to effectively model the AJP process across different operating conditions. 
It learned and leveraged the process-structure relationship in AJP from the source dataset. By utilizing the knowledge transfer method,  
the proposed method could rapidly predict the outcomes of target printing conditions with minimal additional data.

Several studies have focused on optimizing the process parameters to improve the quality of printed lines. 
Zhang \textit{et al.} utilized Gaussian process regression in their study, while another study conducted by the same authors employed response surface methodology to model the intricate relationships between key process parameters, such as shield gas flow rate, center gas flow rate, and print speed, and their impact on structural features, including line width, thickness, and edge roughness \cite{zhang2020multi, zhang2020hybrid}.
Zhang \textit{et al.}  extended an ML-based optimization approach focusing on improving droplet morphology \cite{zhang2023machine}. 
The approach combines various ML techniques, including K-means clustering for identifying cause-effect relationships, SVMs for defining an optimal operating window, and Gaussian process regression for process modeling. 
Their work shows the ability to control the deposited droplet diameter and thickness to achieve consistent, high-quality printing outcomes.
By leveraging these ML approaches, they demonstrated the predictive capabilities and found the optimal process parameters for enhancing the quality of the printed outcomes while minimizing the experimental trials.
These ML-based process-structure modeling approaches have significantly contributed to enhancing the ability to achieve accurate and consistent structural control in AJP manufacturing through both predictive modeling and optimization strategies.

The structural characteristics of printed outcomes are closely linked to their functional performance. 
ML-based structure-property modeling approaches in AJP have thus focused on defining the relationship between these structural features and the resulting properties. 
Salary \textit{et al.} aimed to utilize visual in-situ monitoring data to extract meaningful structural features using a shape-from-shading algorithm \cite{salary2017online}. 
They correlated the extracted feature, in this case, the cross-sectional profile, with the electrical resistance as a functional property. 
Their proposed framework enabled property prediction based on structural characteristics obtained from real-time monitoring.
Similarly, Sun \textit{et al.} developed an image-based quality predictive modeling approach \cite{sun2017quality}. 
Their method leverages a principal component analysis method to extract the features from the microscopic images, capturing the structural characteristics and intensity distribution of printed lines. 
Then, utilizing regression models, this approach found the quantitative relationship between the morphological characteristics and electrical resistance properties, while predicting the overspray. 
This enables multiproperty predictions that consider interdependencies between structures and properties.
Salary \textit{et al.} introduced a sparse representation classification framework that accurately predicts and classifies electrical resistance by correlating it with in-situ monitored line morphology \cite{salary2020sparse} .
The development of such ML-driven structure-property relationship modeling approaches has enhanced the prediction of the performance of printed results based on structural characteristics, enabling more targeted approaches for quality control in AJP manufacturing.

The ML-based PSP relationship modeling approaches represent a comprehensive framework for understanding and controlling AJP manufacturing.
Lombardi \textit{et al.} implemented a system that integrates image-based in-situ monitoring and closed-loop control within the AJP manufacturing process \cite{lombardi2019image}. 
Their study demonstrates that the PSP modeling approach enhances the consistency and quality of the printed structure, leading to improved electrical performance. 
By tracking the printed linewidth in real-time, a regression model in the control module continuously adjusts the printing speed, enabling automated optimization of the printed line structure and improving its functional properties.
Meenakshisundaram \textit{et al.}  developed a framework that combines a finite element method with a neural network-biased genetic algorithm and unsupervised ML to optimize capacitive device manufacturing  \cite{meenakshisundaram2021insights}. 
Their study established PSP links by connecting particle size and the microstructural arrangement of particles to predict capacitance variance . 
This approach enables a better management of process parameters, structural development, and final device properties, achieving enhanced control over capacitive performance through integrated PSP modeling.
These comprehensive PSP approaches represent a significant advancement over simple relationship models (process-structure or structure-process only), as they capture holistic feature relationships in AJP. 
This allows for more precise control and optimization of AJP processes and ultimately improves the performance of the printed outcomes.

It is necessary to analyze PSP data from multiple sources to effectively assess process stability and predict manufacturability as individual sensing systems provide only limited insights into AM processes \cite{akhavan2022sensory}.
Studies have shown that fusing multiscale and multimodal data from various sensing systems can enhance the understanding of complex PSP relationships in AM. 
Yang \textit{et al.}  proposed a fusion framework for AM data from multiple builds and simulations \cite{yang2024holistic}. 
This study demonstrated the method's capability for process monitoring, control, establishing causal PSP relationships, and AM qualification. Chen et al. proposed a framework that spatially and temporally fuses multimodal AM data to create a 3D printed part, allowing location-specific quality monitoring \cite{chen2023multisensor}.

\section{Problem Statement}
\label{State}
\begin{figure*}[t]
  \vspace*{4pt}
  \centerline{\includegraphics[width=0.85\textwidth]{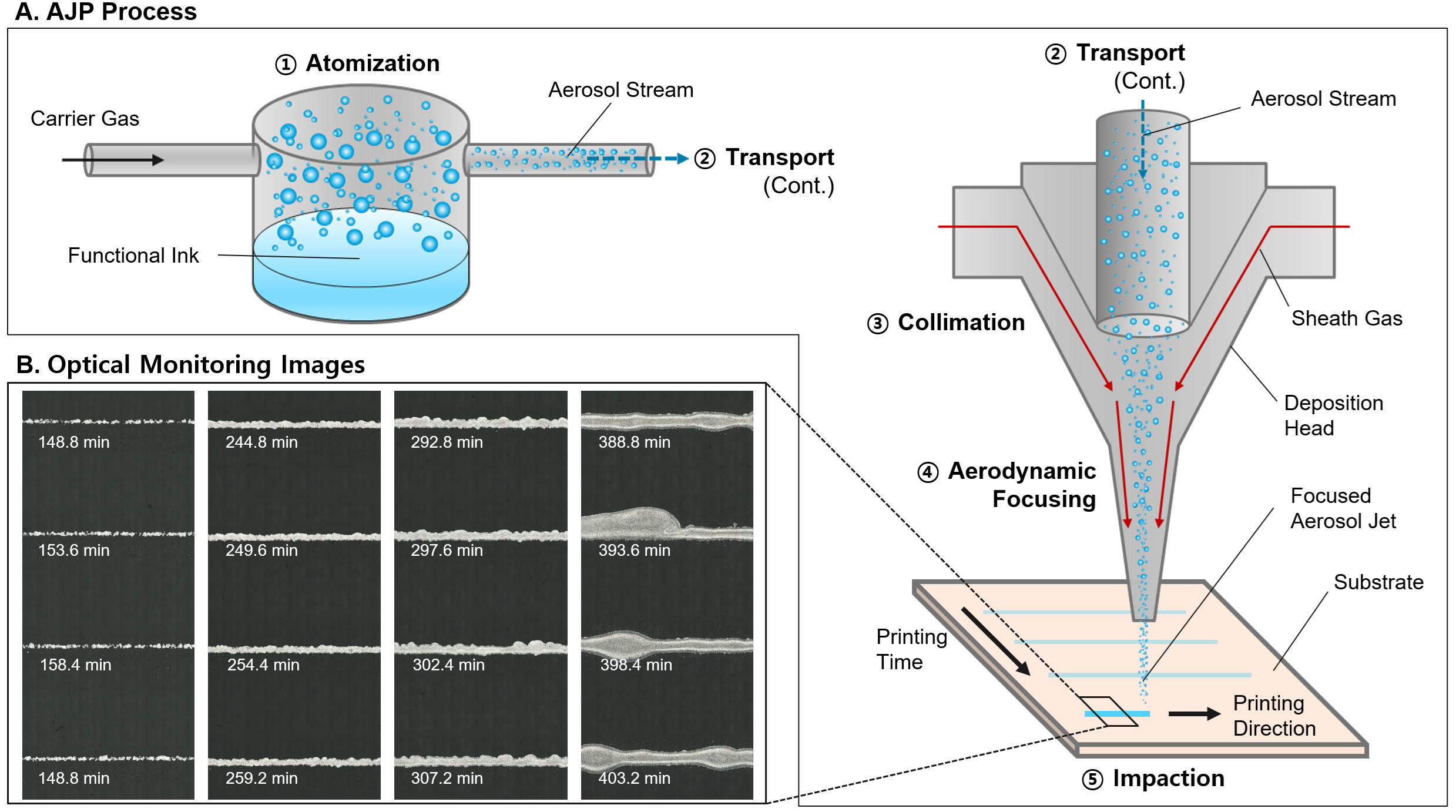}}
    \caption{The Schematic Diagram of the Key Stages in the AJP Process and the  Real Example of Optical Monitoring Images of Printed Lines in Aerosol Jet Printing. (A) This subfigure illustrates the five main stages involved in the AJP process: aerosol droplet generation (atomization), aerosol transport, aerosol beam collimation, aerodynamic focusing, and impaction; (B) This subfigure shows the examples of the post-print optical monitoring images of printed lines at different time points during the AJP process. Each of the four sample images contains four lines, printed sequentially from top to bottom. The top line in each image was printed earlier, while the subsequent lines show gradual increases in width, indicating variations in line morphology over time despite consistent process parameters \cite{yoo2022optical}.}
    \label{fig_Prob}
\end{figure*}

AJP is composed of multiple stages, each containing various mechanisms that interact and connect with one another to form a cohesive process. This interconnected nature of stages and mechanisms underscores the inherent complexities and uncertainties in AJP. 
Figure \ref{fig_Prob} demonstrates the uncertainties and complexities of the AJP process with examples of optical monitoring images of printed lines. Despite maintaining the same process parameters, the morphologies of the printed lines vary depending on the printing time, with sequentially printed lines increasing the line width. These variations are challenging to interpret with current ML approaches. 

More specifically, our study addresses the fusion of multimodal, multiscale data on printed lines' width, shape, height, and density to improve the understanding of their relationships with dynamic changes.
In AJP, the intricate geometries and density of parts necessitate a nuanced understanding of the complex, dynamic physical phenomena involved. 
So far, however, analyzing each modality individually has provided limited insight into these relationships.

Due to the multimodality and complexity of AJP data, achieving precise data fusion is challenging. 
Generative deep learning, particularly denoising diffusion modeling, has demonstrated exceptional performance in capturing and synthesizing multimodal data by learning their complex distributions, over other generative approaches, such as generative adversarial networks \cite{ho2020denoising, goodfellow2020generative}.
Motivated by these capabilities and the unique advantages of diffusion modeling, our study applies a data fusion method based on diffusion modeling, specifically adapting it to AJP \cite{zhao2023ddfm}.

\section{AJP Data Fusion}
\label{Fusion}

\begin{figure*}[t]
    \vspace*{4pt}
    \centerline{\includegraphics[width=0.9\textwidth]{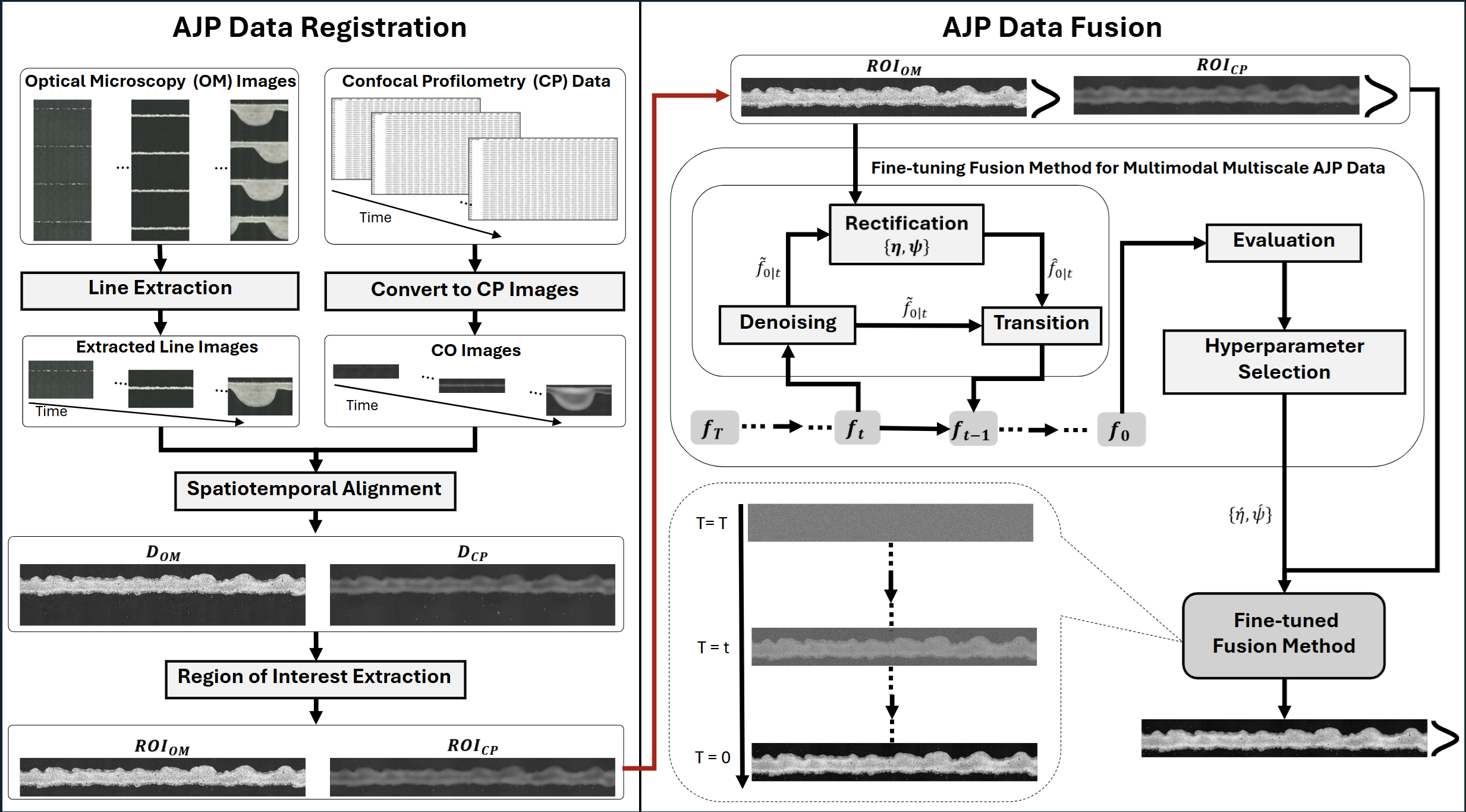}}
    \caption{An Overall Framework for Multimodal Multiscale AJP PSP Fusion.}
    \label{fig_Framework}
\end{figure*}

This section introduces a diffusion-based PSP data fusion methodology designed to 
predicts PSP causality and generates synthesized, fused PSP
features learned from AJP data of  optical microscopy (OM) images and
confocal profilometry (CP) data \cite{zhao2023ddfm, song2020denoising, yoo2022optical}. 
The framework consists of two steps: (1) the spatial and
temporal extraction and registration of PSP features of interest,
utilizing OM and CP data, and (2) data
fusion, predicting causality between the PSP features and
generating synthesized, fused PSP features. Figure \ref{fig_Framework} presents an overview of the proposed fusion framework.  

\subsection{AJP Data Acquisition}
This study utilizes a dataset obtained from experiments conducted at the Air Force Research Laboratory \cite{yoo2022optical}. 
The dataset was created to map both the drift in morphology and the electrical performance of AJP lines over a 16-hour print duration.
The experiment used a mixture of Clariant Prelect TPS 50 G2 silver nanoparticle ink, ethylene glycol, and deionized water in a 0.6:0.3:0.2 volumetric ratio. Deposition was achieved using an Optomec AJ 300-UP Aerosol Jet Deposition System with a Sprint Series Ultrasonic Atomizer MAX and a 150 \textmu{}m nozzle.

The dataset includes OM images and CP data collected using a Keyence VK-X200 optical microscope. OM images, captured at 150x magnification, provide high-resolution two-dimensional (2D) surface morphology and density of 350 \textmu{}m long printed lines at different time points. These images reveal essential details such as line width and deposition uniformity, with a resolution of 50 \textmu{}m to 267 pixels. CP data, also collected with the same microscope, offers high z-height resolution, providing 3D surface height profiles. This data captures topographical variations essential for assessing the structural integrity of the printed lines.

\subsection{AJP Data Registration}
To effectively manage the multimodal data from AJP, we implement a data registration process that aligns various data types, ensuring consistency and accuracy across spatial and temporal domains. This approach is crucial to facilitating our fusion process, which deduces reliable and meaningful insights for understanding the PSP relationship in AJP processes.
This study employs two types of data—the OM images and CP data of the AJP process—to demonstrate the operation of the proposed methodology while other data from the AJP process could also be used, suggesting its potential for broader applications.
We carried out a multistep data registration process to integrate the OM and CP data within our framework. The steps involved are as follows: (1) Extracting lines from the OM images, (2) Converting CP data into CP images, (3) Performing spatiotemporal alignment, and (4) Extracting regions of interest.

In the initial step of data registration, this study performed segmentation on the OM images. 
Each OM image in the dataset contains four printed lines, and the study divided the images so that each resulting image corresponds to a single line. 
Rather than narrowing down the region of interest to just the line itself, we opted to retain the surrounding area in each divided image. 
This approach preserves additional contextual information near the lines, which is crucial for accurate spatial alignment with CP data in the later stage of the data registration process. 
Additionally, the dataset provides temporal information, allowing us to arrange the segmented OM images in their correct chronological order, which is critical for the overall spatiotemporal alignment.

The data registration process also includes converting the CP data into CP images and creating a 2D spatiotemporal representation. 
This enables the use of a common modality between datasets.
This conversion involves projecting the numerical height information onto a 2D surface in a spatiotemporal manner, generating CP images that represent the topographical features of the printed lines.
The study implemented a standardized scaling procedure to ensure consistency across the converted CP images.
We examined the range of height value distribution across all CP data and applied the uniform scale for pixel value assignment in the resulting CP images.
Leveraging the temporal information of CP data from the dataset, the study organized CP images chronologically. 

Once both the OM and CP images are prepared in comparable formats, spatiotemporal alignment is performed to ensure consistency between the datasets. Spatial alignment ensures that the features in the CP images correspond accurately to those in the OM images, maintaining a proper spatial correspondence. 
To achieve this, this step measures the similarity between corresponding regions of the CP and OM images. 
By calculating the correlation between pixel values, the study determines the degree of alignment between the two datasets. 
The cross-correlation function identifies the spatial offset that maximizes the similarity, thereby finding the optimal alignment. 
This ensures precise alignment between the temporally aligned CP and OM images.

\begin{figure*}[t]
\vspace*{4pt}
\centerline{\includegraphics[width=0.85\textwidth]{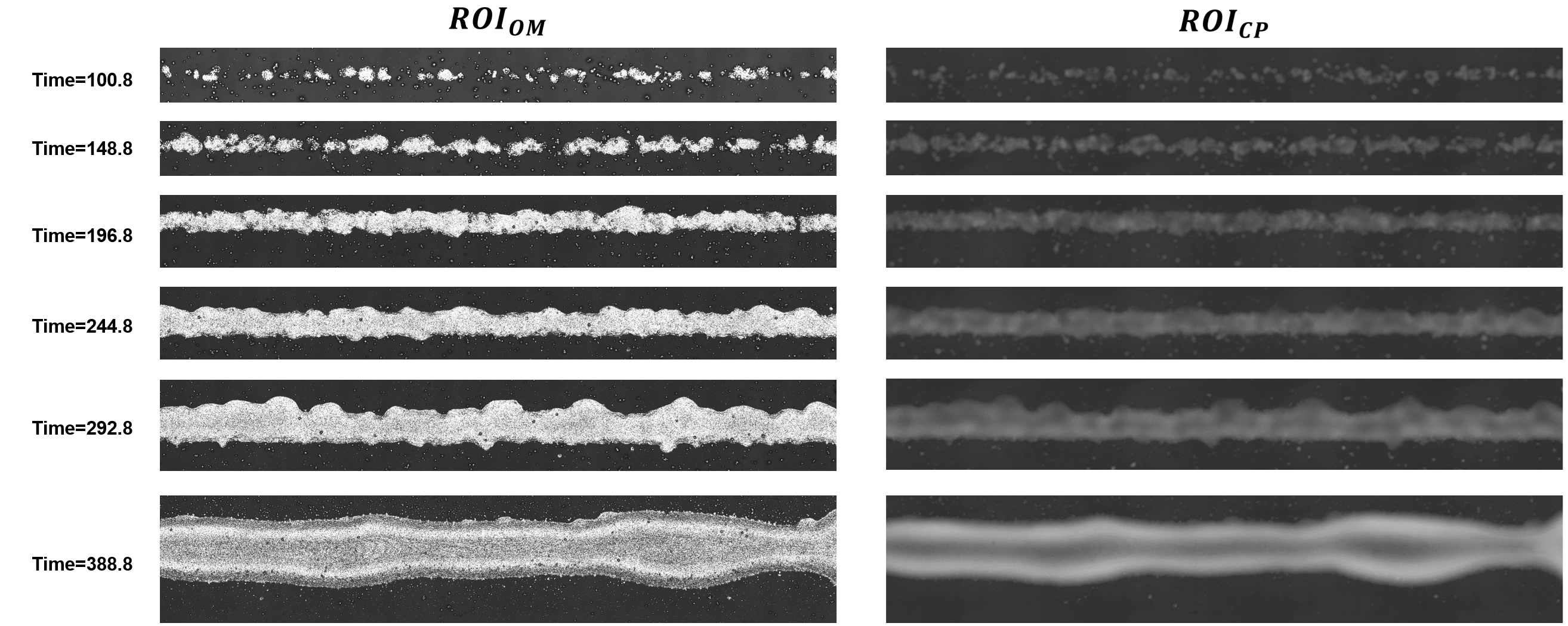}}
    \caption{Example of Registered and Extracted \( \text{ROI}_{\text{OM}}\) and \( \text{ROI}_{\text{CP}}\)Images for Data Fusion. This figure shows an example of registered and extracted \( \text{ROI}_{\text{OM}}\) and \( \text{ROI}_{\text{CP}}\)images, demonstrating both spatiotemporal alignments achieved during the data registration process. This alignment ensures consistency between the two data types helps to form a reliable basis for the subsequent data fusion process.}
    \label{fig_Registered data}
\end{figure*}

After aligning the datasets spatially and temporally, the methodology extracts regions of interest (ROIs) from each aligned dataset. 
This step focuses on isolating the specific portions of the aligned data that are most relevant for analysis. 
In this study, the ROIs correspond to the printed line areas in both the OM and CP datasets. 
Let \( D_{\text{OM}} \) and \( D_{\text{CP}} \) denote the OM and CP data, respectively. Following alignment, regions of interest \( \text{ROI}_{\text{OM}} \subset D_{\text{OM}} \) and \( \text{ROI}_{\text{CP}} \subset D_{\text{CP}} \) are selected from each dataset. 
These ROIs serve as inputs to the subsequent fusion and analysis stages, allowing for the integration of multimodal features extracted from each data type.
By extracting these regions, the methodology narrows the scope of the analysis to the printed features, excluding irrelevant surrounding areas. 
This targeted approach ensures that our analysis is focused on the critical PSP ROI. 
Figure \ref{fig_Registered data} shows the example of \( \text{ROI}_{\text{OM}}\) and \( \text{ROI}_{\text{CP}}\) the of the registered \( D_{\text{OM}} \) and \( D_{\text{CP}} \) used in our study.

\subsection{Generative Data Fusion}
This section presents the denoising diffusion-based fusion
method. The goal of diffusion-based generative modeling is to capture and preserve multimodal AJP PSP features and generate new synthesized fused features, enabling an understanding of the intricate and dynamic nature of PSP linkages in a spatially and temporally fused manner. 

Figure \ref{fig_Framework} illustrates the architecture of AJP PSP fusion. The inputs of the fusion method are pairs of \( \text{ROI}_{\text{OM}}\) and \( \text{ROI}_{\text{CP}}\) and the output is a set of fused PSP features. Diffusion methods generate samples by reversing a forward process in which a complex distribution of initial data is gradually transformed into a pure Gaussian noise distribution via $T$ steps, $\{1,...,T\}$, where $T \in \mathbb{Z}^+$ \cite{ho2020denoising}. The forward process starts with the initial state of data without noise, denoted as $f_0$. 
At each step $t$, where $t \in \mathbb{Z}^+ \cap [0, T]$, a method gradually adds Gaussian noise to the data, $f_t$, the noisy data, at the step. 
Each step of the forward process can be defined by Equation \ref{Eq_Forward}:
\begin{equation}\label{Eq_Forward}
    q(f_t|f_{t-1}) = N(f_t; \sqrt{1-\beta_t}f_{t-1}, \beta_tI)
\end{equation}
, where $\beta$ is the variance. The approximate posterior distribution of the forward process can be defined as Equation \ref{Eq_Forward_Approximate}:
\begin{equation}\label{Eq_Forward_Approximate}
    q(f_{1:T}|f_0) = \prod_{t=1}^{T} q(f_t|f_{t-1})
\end{equation}

Leveraging the forward process, the reverse process learns the distribution of the data by gradually denoising, starting with the pure noise data, $f_T$, through a Markovian process. This process at step $t$ can be defined as Equation \ref{Eq_Reverse}:

\begin{align}\label{Eq_Reverse}
    p(f_{t-1} | f_{t}, ROI_{OM}, ROI_{CP}) = 
    N\big(f_{t-1}; & \ \mu(f_t, t, ROI_{OM}, ROI_{CP}), \nonumber \\
    & \ \Sigma(f_t, t, ROI_{OM}, ROI_{CP}) \big)
\end{align}
, where $\mu$ is the mean. 
The entire reverse process with learned Gaussian noise then becomes Equation \ref{Eq_Reverse_Entire}:

\begin{equation}\label{Eq_Reverse_Entire}
    p_(f_{0:T}) = p(f_T)\prod_{t=1}^{T} p_(f_{t-1}|f_t, ROI_{OM}, ROI_{CP})
\end{equation}

\begin{figure*}[tbp]
    \vspace*{4pt}
    \centerline{\includegraphics[width=0.85\textwidth]{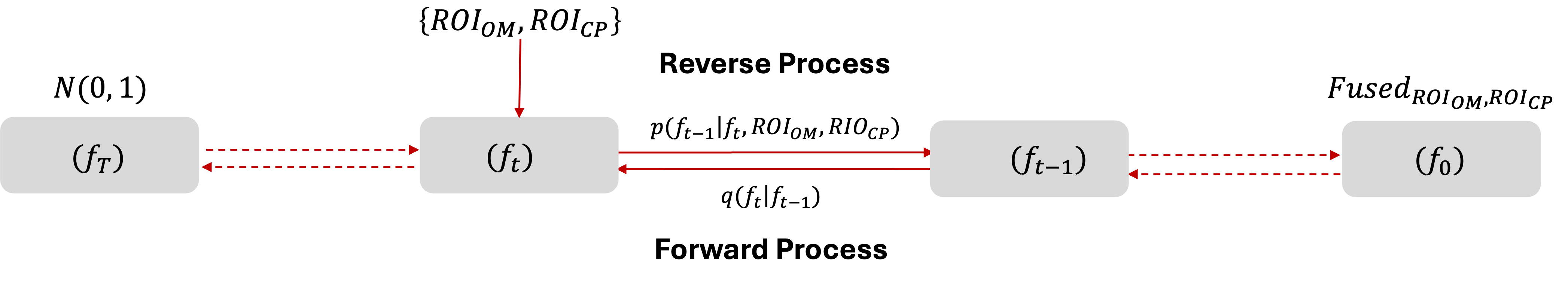}}
    \caption{A Denoising Diffusion-Based Fusion of \( \text{ROI}_{\text{OM}}\) and \( \text{ROI}_{\text{CP}}\)  }
    \label{ddpm}
\end{figure*}

Figure \ref{ddpm} indicates the forward and reverse processes of fusing \( \text{ROI}_{\text{OM}}\) and \( \text{ROI}_{\text{CP}}\) via a diffusion modeling approach.

The proposed method leverages non-Markovian implicit diffusion modeling, which has the advantage of not requiring the simulation of a Markov chain for many steps to produce a sample \cite{song2020denoising}. 
The fusion method contains three main steps. The first step is the denoising step. Through denoising, the method receives fused AJP \( \text{ROI}_{\text{OM}}\) and \( \text{ROI}_{\text{CP}}\)  from step $t+1$, $f_t$, and predicts an initial denoised data, $\tilde f_{0|t}$ as shown in Equation \ref{Eq_denoising}:

\begin{equation}\label{Eq_denoising}
    \tilde f_{0|t}= \frac{1}{\sqrt{\Bar{\alpha_t}}}(f_t+(1-\Bar{\alpha_t}s(t))
\end{equation}
The second is the rectification, where the method enhances the preservation of cross-modality information from \( \text{ROI}_{\text{OM}}\) and \( \text{ROI}_{\text{CP}}\). It rectifies the preliminary fused data $\tilde f_{0|t}$ with \( \text{ROI}_{\text{OM}}\) and \( \text{ROI}_{\text{CP}}\) and predicts $\hat{f}_{0|t}$ as seen in Equation \ref{Eq_PSP_fusion_estimation}:

\begin{equation}\label{Eq_PSP_fusion_estimation}
    \hat{f}_{0|t} = x + ROI\end{equation}
, where $x = \tilde{f}_{0|t} - ROI$.
The third step is the transition step. The transition predicts less noisy fused AJP \( \text{ROI}_{\text{OM}}\) and \( \text{ROI}_{\text{CP}}\), $f_{t-1}$ as shown in Equation \ref{DDIM}:
\begin{equation}\label{DDIM}
       f_{t-1}=\frac{\sqrt{\alpha_t}(1-\Bar{\alpha}_{t-1})}{1-\Bar{\alpha_t}}f_t+\frac{\sqrt{\Bar{\alpha}_{t-1}}\beta_t}{1-\Bar{\alpha_t}} \hat{f}_{0|t}
\end{equation}
, where $\alpha_t=1-\beta_t$, $\Bar{\alpha_t}=\prod_{s=1}^{t} \alpha_s$, and $s$ is a score function \cite{song2020score}. More details of the methodology can be found in \cite{zhao2023ddfm, song2020denoising}.

This study adopted a fine-tuning algorithm to optimize two hyperparameters $\eta$ and $\psi$ of the rectification step. The hyperparameters indicate the preservation of cross-modality information from multimodal features \cite{ zhao2023ddfm}. In particular, fine-tuning ensures that the pre-trained fusion model accurately captures and integrates the unique characteristics and dependencies within AJP PSP data by identifying the optimal values of $\eta$ and $\psi$. This maximizes the effective fusion of AJP PSP features, ensuring the preservation of both the original features and the causal PSP relationships. More specifically, the hyperparameters control the balance between preserving the distinct characteristics of each modality and ensuring smooth fusion \cite{ zhao2023ddfm}.  Therefore, fine-tuning these hyperparameters ensures optimal, customized fusion results by maintaining cross-modality consistency and minimizing fusion artifacts unrelated to the original features. 

Algorithm \ref{alg:diffusion_fine_tuning} indicates the details of fine-tuning for the hyperparameters of the rectification step. The inputs for the fine-tuning step are pairs of $\{ \text{ROI}_{\text{OM}}, \text{ROI}_{\text{CP}} \}$
and $\{\eta, \psi\}$. We utilized 30 extracted $\{ \text{ROI}_{\text{OM}}, \text{ROI}_{\text{CP}} \}$
 pairs of images. We set the number of denoising steps as $T=100$. Later, to evaluate the results, we calculated the average values of structural similarity index measure (SSIM) at each fine-tuning iteration \cite{ma2019infrared}.  SSIM consists of three components: loss of correlation and luminance and contrast distortion as illustrated in Equation \ref{ssim_f}:

\begin{equation}
\label{ssim_f}
SSIM_{l,f} = \sum_{l,f} \left( \frac{2\mu_l \mu_f }{\mu_l^2 + \mu_f^2 } \right) 
\left( \frac{2\sigma_l \sigma_f }{\sigma_l^2 + \sigma_f^2} \right) 
\left( \frac{\sigma_{lf} }{\sigma_l \sigma_f} \right)
\end{equation}
, where $l$ and $f$ denote the image patches of source and fused images in a sliding window, respectively; $\sigma_{lf}$ denotes the covariance; $\sigma_l$ and $\sigma_f$ denote the standard deviation; and $\mu_l$ and $\mu_f$ denote the mean values of source and fused images [ref6]. the SSIM between pairs of $\{ \text{ROI}_{\text{OM}}, \text{ROI}_{\text{CP}} \}$
 and their fused image, $f$, can be written as in Equation \ref{SSIM}:

\begin{equation}
\label{SSIM}
SSIM = SSIM_{\text{ROI}_{\text{OM}},f} + SSIM_{\text{ROI}_{\text{CP}},f} 
\end{equation}

We calculate the average SSIM values for each hyperparameter value as in Equation \ref{ave}:
\begin{equation}\label{ave}
\overline{\text{SSIM}} = \frac{\sum_{i=1}^{N} \text{SSIM}_i}{N}
\end{equation}
, where $N$ is the number of samples.
Finally, the hyperparameters with the optimum values are chosen for the final fusion step utilizing Equation \ref{fusion}:

\begin{equation}\label{fusion}
\text{Fused}_{i} = \text{Fusion}_{\text{fine-tuned}}(\text{ROI}_{\text{OM},i}, \text{ROI}_{\text{CP},i}; \eta^*, \psi^*), \quad i = 1, \ldots, N
\end{equation}

\begin{algorithm}[t]
\caption{Fine-Tuning}
\label{alg:diffusion_fine_tuning}
\begin{algorithmic}[1]
\STATE \textbf{INPUT:} Pairs of $\{ \text{ROI}_{\text{OM}}, \text{ROI}_{\text{CP}} \}_{i=1}^{30}$, hyperparameters $\eta$ and $\psi$, and $T=1, \dots, 100$
\STATE \textbf{OUTPUT:} Optimal hyperparameters $(\eta^*, \psi^*)$

\STATE \textbf{Step 1: Fine-tune $\eta$ with $\psi$ constant}
\STATE Initialize $\psi = \psi_0$
\STATE Define $\eta$ range as $\text{uniform from 0 to 100}$
\FOR{each candidate value $\eta_i$ in the range}
    \FOR{each image pair $(\text{ROI}_{\text{OM}, j}, \text{ROI}_{\text{CP}, j})$ in $\{ \text{ROI}_{\text{OM}}, \text{ROI}_{\text{CP}} \}_{i=1}^{30}$}
        \FOR{$t = T, T-1, \dots, 1$}
            \STATE Compute Equations \ref{Eq_denoising} to \ref{DDIM}
        \ENDFOR
    \ENDFOR
    \STATE Compute SSIM using Equation \ref{SSIM}
\ENDFOR
\STATE Compute average SSIM using Equation \ref{ave}
\STATE Select $\eta^* = \arg \max_{\eta_i} \text{average SSIM}$

\STATE \textbf{Step 2: Fine-tune $\psi$ with $\eta^*$ constant}
\STATE Set $\eta = \eta^*$
\STATE Define $\psi$ range as $\text{uniform from 0 to 100}$
\FOR{each candidate value $\psi_k$ in the range}
    \FOR{each image pair $(\text{ROI}_{\text{OM}, j}, \text{ROI}_{\text{CP}, j})$ in $\{ \text{ROI}_{\text{OM}}, \text{ROI}_{\text{CP}} \}_{i=1}^{30}$}
        \FOR{$t = T, T-1, \dots, 1$}
            \STATE Compute Equations \ref{Eq_denoising} to \ref{DDIM}
        \ENDFOR
    \ENDFOR
    \STATE Compute SSIM using Equation \ref{SSIM}
\ENDFOR
\STATE Compute average SSIM using Equation \ref{ave}
\STATE Select $\psi^* = \arg \max_{\psi_k} \text{average SSIM}$

\RETURN $(\eta^*, \psi^*)$
\end{algorithmic}
\end{algorithm}

\section{Results}
\label{results}
\subsection{Fine-Tuning}
\begin{figure}[t]
    \vspace*{4pt}
    \centering
    \begin{subfigure}[t]{\columnwidth}
        \includegraphics[width=0.85\columnwidth]{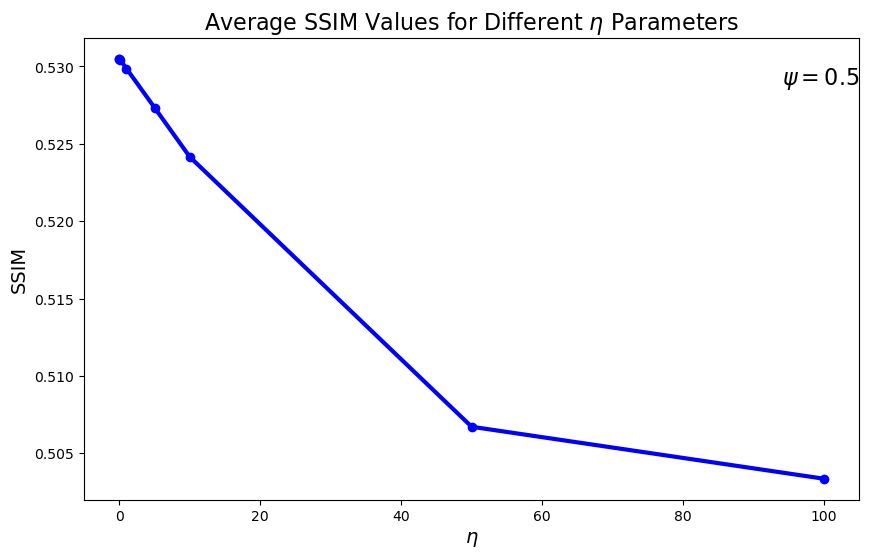}
        \label{fig:sub1}
    \end{subfigure}
    \vspace{-5pt}
    
    \begin{subfigure}[t]{\columnwidth}
        \includegraphics[width=0.85\columnwidth]{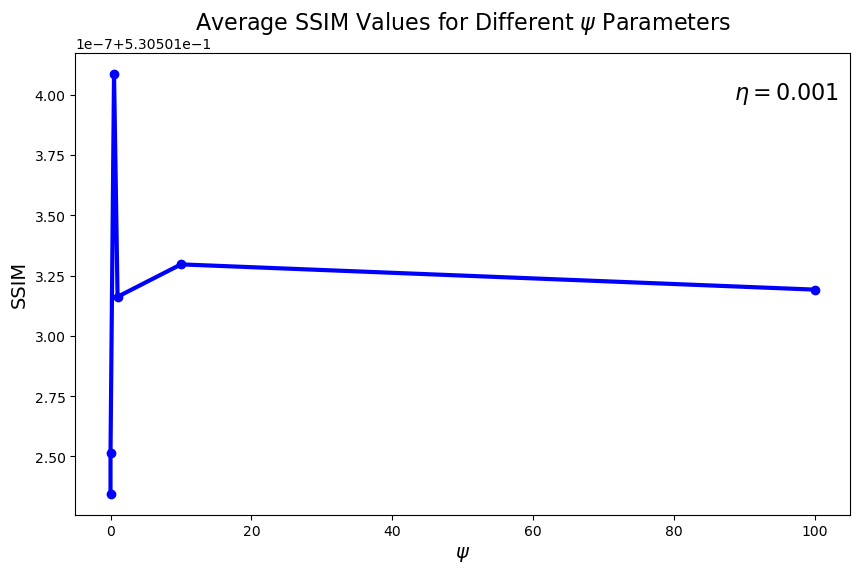}
        \label{fig:sub2}
    \end{subfigure}
    
    \caption{Fine-Tuning Results for Hyperparameters $\eta$ and $\psi$.}
    \label{fine}
\end{figure}

\begin{figure}[t]
    \vspace*{4pt}
    \centerline{\includegraphics[width=0.85\columnwidth]{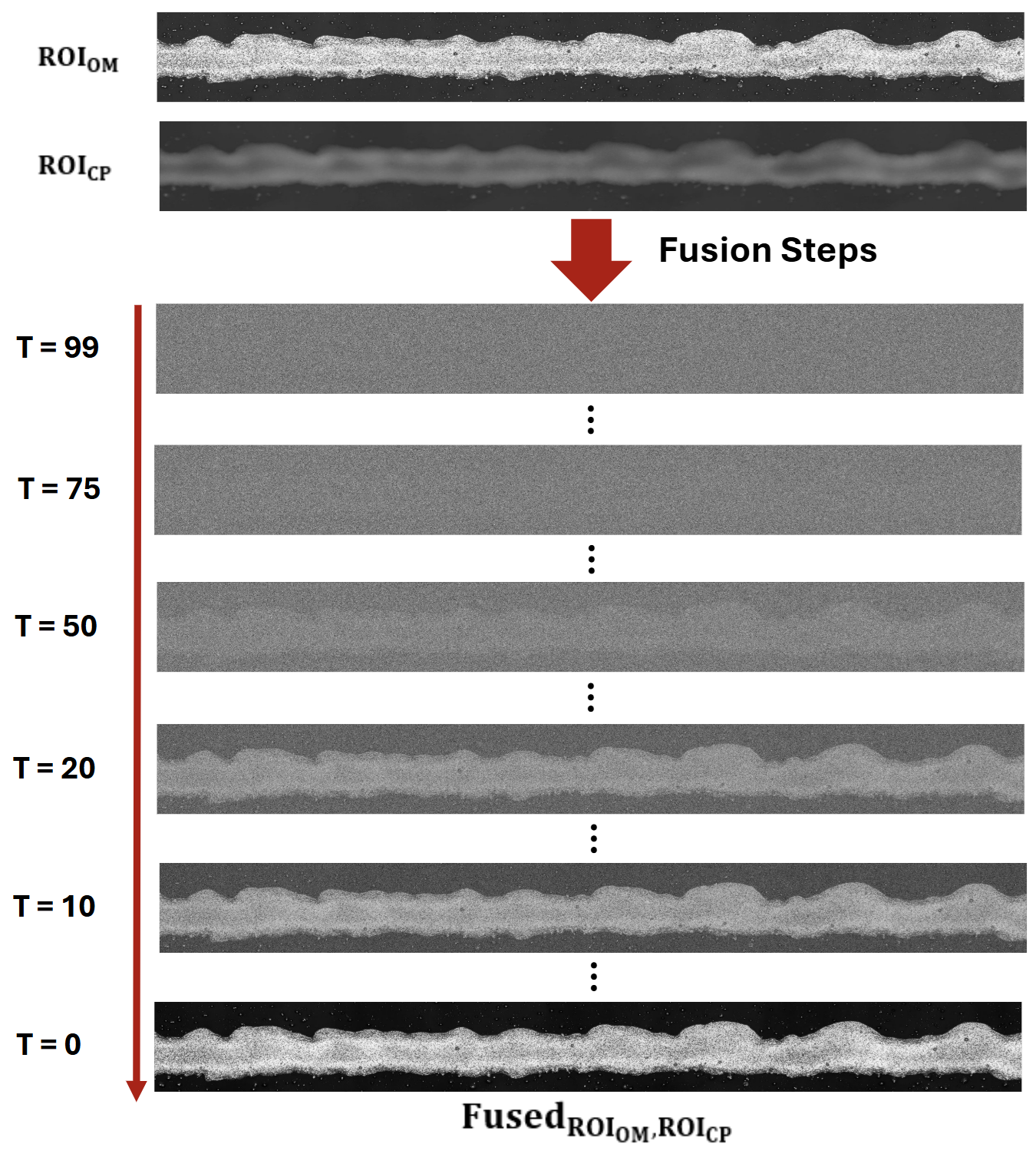}}
    \caption{An Illustration of  Fusion Steps For \( \text{ROI}_{\text{OM}}\) and \( \text{ROI}_{\text{CP}}\) Via The Proposed Fusion Method.}
    \label{Fused}
\end{figure}

Figure \ref{fine} displays the fine-tuning results for the hyperparameters $\eta$ and $\psi$. We calculated the average ssim values for the pairs of $\{\text{ROI}_{\text{OM}}, \text{ROI}_{\text{CP}}\}$
 images using Equation \ref{ave}. We selected the optimum hyperparameter values based on the highest SSIM value. Later, we utilized the optimum hyperparameter values for the final fusion task using Equation \ref{fusion}. 
 Figure \ref{Fused} represents the fusion steps of a pair of $\{\text{ROI}_{\text{OM}}, \text{ROI}_{\text{CP}}\}$
 images, starting from the noisy fused image at step $T=99$ to the final fused images at step $T=0$. The fused image captures and predicts original features from both $\{\text{ROI}_{\text{OM}}, \text{ROI}_{\text{CP}}\}$
 images and generates new, synthesized PSP features.

 \subsection{Ablation Study}
 We eliminated the denoising diffusion part from the method and compared its capability in fusing AJP PSP features with the proposed fusion method as an ablation experiment \cite{zhao2023ddfm}. 
 We used the optimal fine-tuned values for the hyperparameters $\eta$ and $\psi$ for the fusion of the pairs of $\{\text{ROI}_{\text{OM}}, \text{ROI}_{\text{CP}}\}$. Later, we calculated the SSIM and average SSIM values as in Equations \ref{SSIM} and \ref{ave}. The calculated average SSIM values are
$0.5305$ and $0.5206$ for the proposed and non-generative fusion method, respectively. 
The comparison specifically focuses on the effect of diffusion-driven denoising in the fusion process. The diffusion-based fusion incorporates an iterative denoising process, where the model progressively refines the fused representation by removing noise. The results indicate
the superiority of the denoising diffusion approach over the non-generative model in fusing AJP PSP features.

\subsection{Discussion}

\begin{figure*}[!t]
    \vspace*{4pt}
    \centerline{\includegraphics[width=0.82\textwidth]{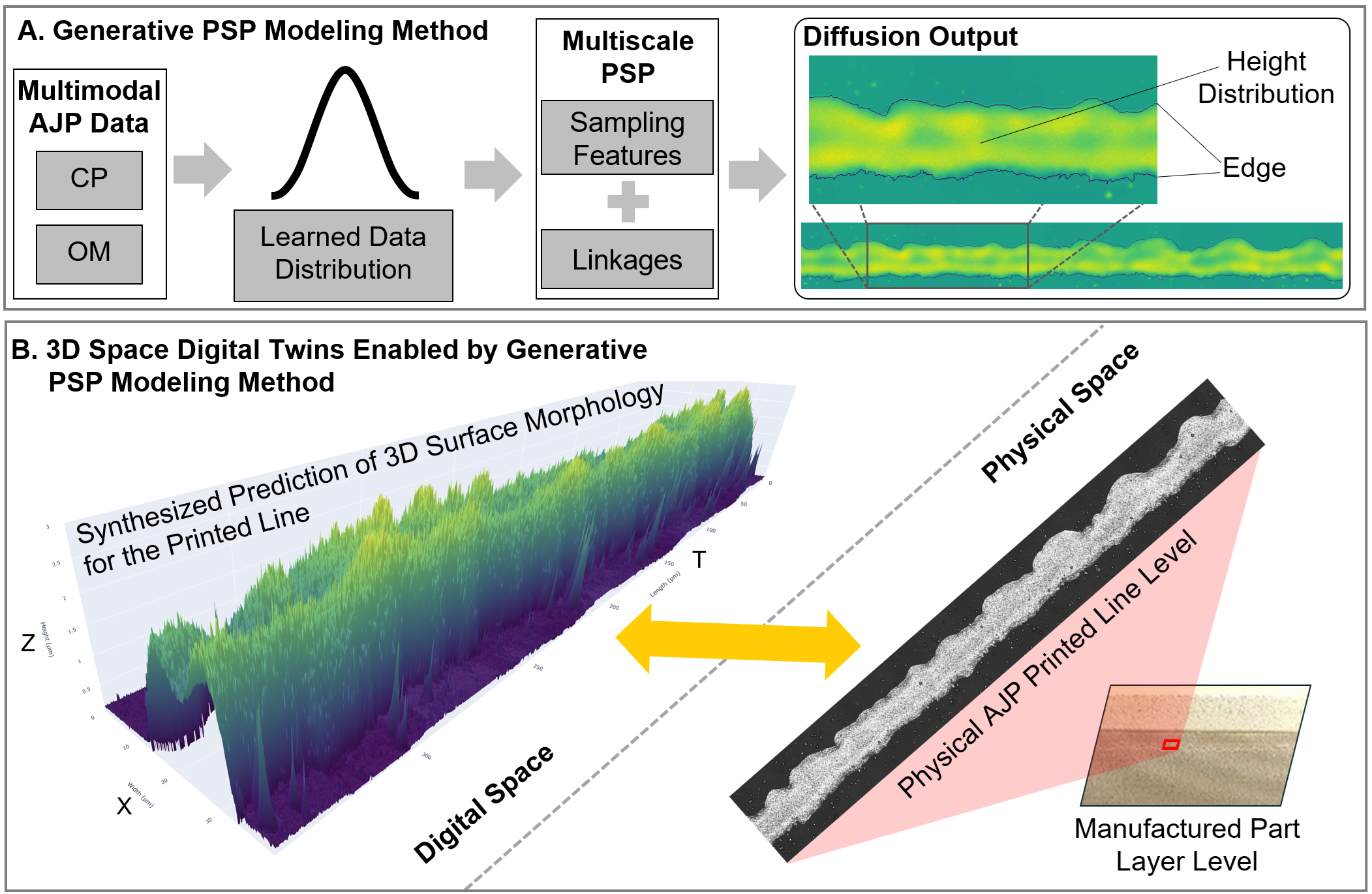}}
    \caption{A Demonstration of DT Construction Enabled by the Proposed Method. (A) The Denoising Diffusion Modeling-Based Fusion Method Learned the Distribution of AJP PSP Features and Predicted 2D Fused Features of \( \text{ROI}_{\text{OM}}\) and \( \text{ROI}_{\text{CP}}\). In the diffusion output, the black line represents the edge, and the color distribution indicates the height distribution of the synthesized PSP feature of the predicted line. (B) DT in 3D space enabled by our proposed method. The synthesized 3D surface morphology serves as a predictive DT corresponding to the physical AJP printed line level of the manufactured part. $Z$, $X$, and $T$ correspond to height, width, and time, respectively.}
    \label{Fig_Flow}
\end{figure*}

The generative learning capability of diffusion model-based data fusion effectively captures and preserves the original information. Furthermore, it generates new, previously unseen data beyond existing sensor measurements by leveraging the learned distribution of AJP PSP data. This provides data-driven insights into potential outcomes under various conditions—insights that traditional simulations or raw sensor data alone cannot provide—as illustrated in Figure~\ref{Fig_Flow}.

The predictive and visualization capabilities for identifying causal linkages offered by the proposed method are invaluable for enhancing data-driven DT construction. As demonstrated in Figure~\ref{Fig_Flow}, the fusion of $\text{ROI}_{\text{OM}}$ and $\text{ROI}_{\text{CP}}$ data synthesizes predicted PSP linkages, enabling improved visualization and real-time spatiotemporal monitoring and analysis. The resulting 3D image effectively illustrates height and width variations over time. Through fusion, the method learns the underlying data distribution and generates features that highlight causal relationships between these geometrical characteristics, revealing previously hidden interconnections while preserving original features. Such an approach allows a predictive and comprehensive exploration of PSP causal relationships in AJP, identifying variations that may significantly influence final manufactured component properties.  To do so, complex graph models can be employed to capture causality in PSP relationships, as they have demonstrated strong capabilities across various domains \cite{sabri2020monitoring, elhambakhsh2021developing, elhambakhsh2022scan, elhambakhsh2023latent}.

Ultimately, the proposed generative framework is expected to enhance control and quality assurance within AJP processes—including real-time monitoring, anomaly detection, and dynamic process planning—by providing new, data-driven generative simulation capabilities based on the newly revealed features.

\section{Concluding Remarks and Future Work}
\label{concl}
This paper introduces a novel methodology based on denoising diffusion modeling for fusing multimodal and multiscale AM data in electronics. The proposed fusion approach integrates geometrical data of AJP parts to effectively combine multimodal features. The methodology is further fine-tuned to accommodate the unique characteristics inherent to AJP processes. The resulting synthesized data captures the intricate PSP relationships within AJP, providing deeper insights into underlying system dynamics.
In future work, we aim to enhance this fusion framework by incorporating a domain adaptation approach that leverages causal relationships among multimodal features as explicit domain knowledge.



\nocite{*}

\bibliographystyle{asmeconf}  


\end{document}